\documentclass[final,3p,twocolumn]{elsarticle}
\usepackage{graphicx}
\journal{Solid State Communications}
\begin{document}
\begin{frontmatter}
\title{Elastic Properties of KH$_2$PO$_4$ at the ferroelectric phase transition }

\author[hppi]{A.E.~Petrova}
\author[hppi]{S.M.~Stishov}
\address[hppi]{Institute for High Pressure Physics , 142190 Troitsk, Moscow, Russia}
\ead{sergei@hppi.troitsk.ru}

\begin{abstract}
We report results of ultrasonic studies of a single crystal of KDP in the temperature range 2-300 K. The longitudinal and transverse sound velocities along [001] and [100] directions were obtained as functions of temperature. The analysis of the temperature evolution of pseudomoduli  $c_{11}$ and $c_{33}$ in the ferroelectric phase shows that these quantities can be approximated by the expression $c_{ii}\sim(T-T_0)^{0.5}$ in the temperature range 0.001K$<T_c-T<$0.2K.
\end{abstract}

\end{frontmatter}

\section{Introduction}
As is widely known, potassium dihydrogen phosphate (KH$_2$PO$_4$ or KDP) experiences a phase transition to a ferroelectric state when cooled below $\sim$122K~\cite{Iona1962, Strukov1998}. At the phase transition the crystal structure of KDP changes from a tetragonal (\={4}2m) to an orthorhombic (2mm) structure due to an ordering of the positions of the hydrogen atoms. The ferroelectric transition in KDP at ambient pressure is often considered as an archetypical example of a first order phase transition close to the second order one, which is not quite true. The point is that although the changes of thermodynamic properties at the phase transition in KDP are indeed very small~\cite{Strukov1998}, the order parameter jumps at the transition by about 1.8 $\mu C/cm^2$ Ref.~\cite{Benepe1971}, which is quite significant when compared to the saturated polarization $P_s \approx 5.1$ $\mu C/cm^2$ Ref.~\cite{Samara1973}. This implies that strong first order features in the phase transition of KDP can not be ignored. A tiny hysteresis of the transition temperature measured during cooling and heating (see later in the paper) reflects the small volume and entropy change across the phase transition. The transition becomes second order at a tricritical point at $\sim$2600 bar, which was well documented in a number of publications~\cite{Schmidt1976, Bastie1978, Aleksandrov1982}. The anomalous behavior of the thermodynamic functions at the phase transition in KDP at ambient pressure was explained in~\cite{Strukov1971, Aleksandrov1982} by the proximity to this tricritical point. At the same time, limited data available on the elastic properties of KDP are usually not taken into account when analyzing its phase diagram.
However, it should be emphasized that elastic properties of ferroelectrics obtained from ultrasound measurements could differ significantly from the ones measured under static conditions. In the present case, propagations of the longitudinal sound waves in the ferroelectric phase of KDP, causing local polarizations in the Z direction, are unable to produce the corresponding $u_{xy}$ mechanical strain,  due to a frequency limitation. As a result the  quantities $c_{ii}$,  calculated through the standard expression  $c_{ii} = V_l^2 \rho$, are, in case of the ferroelectric phase, not true elastic moduli. The quantities analogous to  $c_{33}$ and $c_{11}$  of the paraelectric phase can therefore be called  "pseudomoduli". Though hereafter we will not use different designations for these quantities, their nature will be obvious to the reader from the context of the discussion.     It is also important to remember that according to  Ref.~\cite{Geguzina1967} the propagation of longitudinal ultrasound waves in KDP along the polar axis Z creates an intensive  depolarizing electrical field, which should strongly suppress the corresponding  anomalies at the phase transition. An attempt to verify this statement is part of the current study.

Elastic properties of KDP at the ferroelectric phase transition had attracted considerable attention since Mason's seminal paper~\cite{Mason1946}. In its paraelectric tetragonal phase, KDP is piezoelectric and the $u_{xy}$ mechanical strain is piezoelectrically coupled to the polarization along the Z direction via the piezoelectric constant $d_{36}$. The corresponding elastic modulus $c_{66}$ shows (the Curie-Weiss) anomalous behavior, as follows from shear wave propagation along the [100] direction with [010] polarization~\cite{Garland1969}. These measurements could only be made in the paraelectric phase due to the strong scattering of sound waves at domain boundaries in the ferroelectric phase. Nevertheless the corresponding data were obtained by a Brillouin-scattering technique~\cite{Brody1968, Brody1974}, which revealed the anomalous behavior of $c_{66}$ also in the ferroelectric phase, below $T_c$. Upon transition to the ferroelectric orthorhombic phase three more piezoelectric constants become active and as a result the mechanical strains $u_{xx}$, $u_{yy}$ and $u_{zz}$ couple to the polarization in the Z direction as well.  Consequently, elastic moduli such as $c_{11}$, $c_{22}$ and $c_{33}$ may show anomalous properties in the ferroelectric phase. However, to the best of our knowledge the only attempt to measure longitudinal moduli in KDP was made in Ref.~\cite{Zwicker1946, Harnik1969}, and with limited accuracy. Note that one cannot expect an anomalous behavior of $c_{44}$ and $c_{55}$ since the corresponding piezoelectric constants $d_{34}$ and $d_{35}$ are equal to zero - although high order coupling can not be excluded, which would lead to insignificant variation of $c_{44}$ and $c_{55}$ in the vicinity of the phase transition.

\begin{figure}[htb]
\includegraphics[width=75mm]{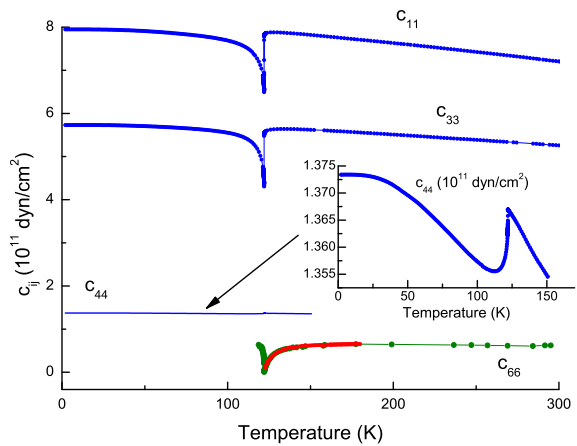}
\caption{\label{fig1} Temperature dependence of elastic moduli  $c_{11}$, $c_{33}$, $c_{44}$ and $c_{66}$ in KDP. Also shown (in red/green {\bf (?)}) is Brillouin scattering data for $c_{66}$ from Ref.~\cite{Brody1968}.}
 \end{figure}
\section{Experimental}
We report results of ultrasonic studies of a single crystal of KDP in the temperature range 2-300K. During the course of this study, several runs were performed using the digital pulse-echo techniques (see~\cite{Petrova2009} and references therein). Several samples of KDP of 1.5-2 mm thicknesses with orientations along [001] and [100] were cut from larger high quality single crystals. The corresponding surfaces of the samples were polished to be parallel and optically flat. Since the elastic and piezoelectric properties of ferroelectric materials depend on electrical boundary conditions, the samples were plated with gold or wrapped with a silver leaf~\cite{Mason1946}.
\begin{figure}[htb]
\includegraphics[width=75mm]{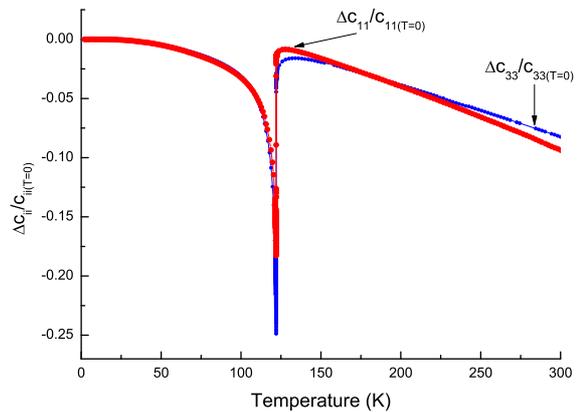}
\caption{\label{fig2} Reduced temperature dependence of the elastic moduli  $c_{11}$ and $c_{33}$.}
 \end{figure}
\begin{figure}[htb]
\includegraphics[width=75mm]{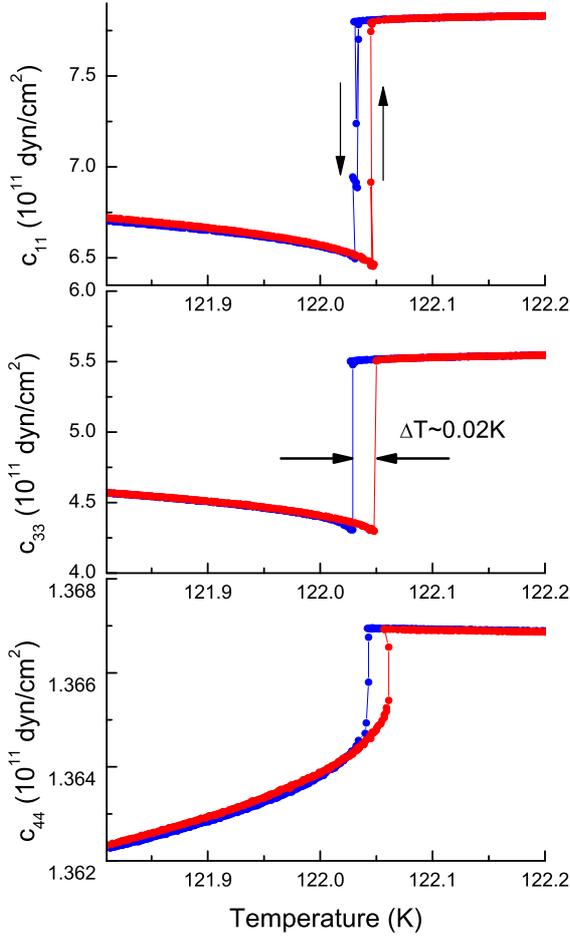}
\caption{\label{fig3} The temperature hysteresis of the elastic moduli $c_{11}$, $c_{33}$, $c_{44}$ at the phase transition in KDP.}
 \end{figure}

The 36$^\circ$ Y ( P-wave) and 41$^\circ$ X (S-wave) cut $LiNbO_3$ transducers were bonded to the samples with silicon grease. For the measurements the experimental setup was placed within the PPMS cryostat, with a temperature control ability of about 0.001K. The elastic moduli $c_{11}$, $c_{33}$, $c_{44}$ and $c_{66}$ were obtained  from the longitudinal and transverse sound velocities along the [001] and [100] directions.  The domain structure does not prevent propagation of longitudinal ultrasound waves along the [001], [100] and [010] directions, nor a shear wave in the [001] direction. Of course, $c_{66}$ could only be measured at $T>T_c$. The results of the experiments are illustrated in Fig.~\ref{fig1}. Shown therein is also literature data, which characterized the behavior of $c_{66}$ in the vicinity of the phase transition.

\begin{figure}[htb]
\includegraphics[width=75mm]{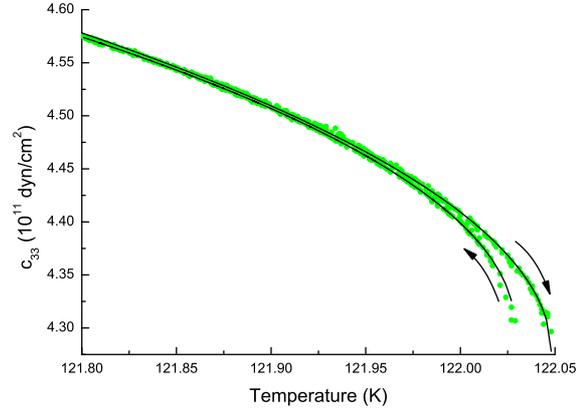}
\caption{\label{fig4} Splitting of the up and down curves of the temperature dependence of the elastic moduli in the hysteresis area. {\bf If you used two different colours for 'up' and 'down', it could be clearer for $T\le 122.00$K.}}
 \end{figure}
\begin{figure}[htb]
\includegraphics[width=75mm]{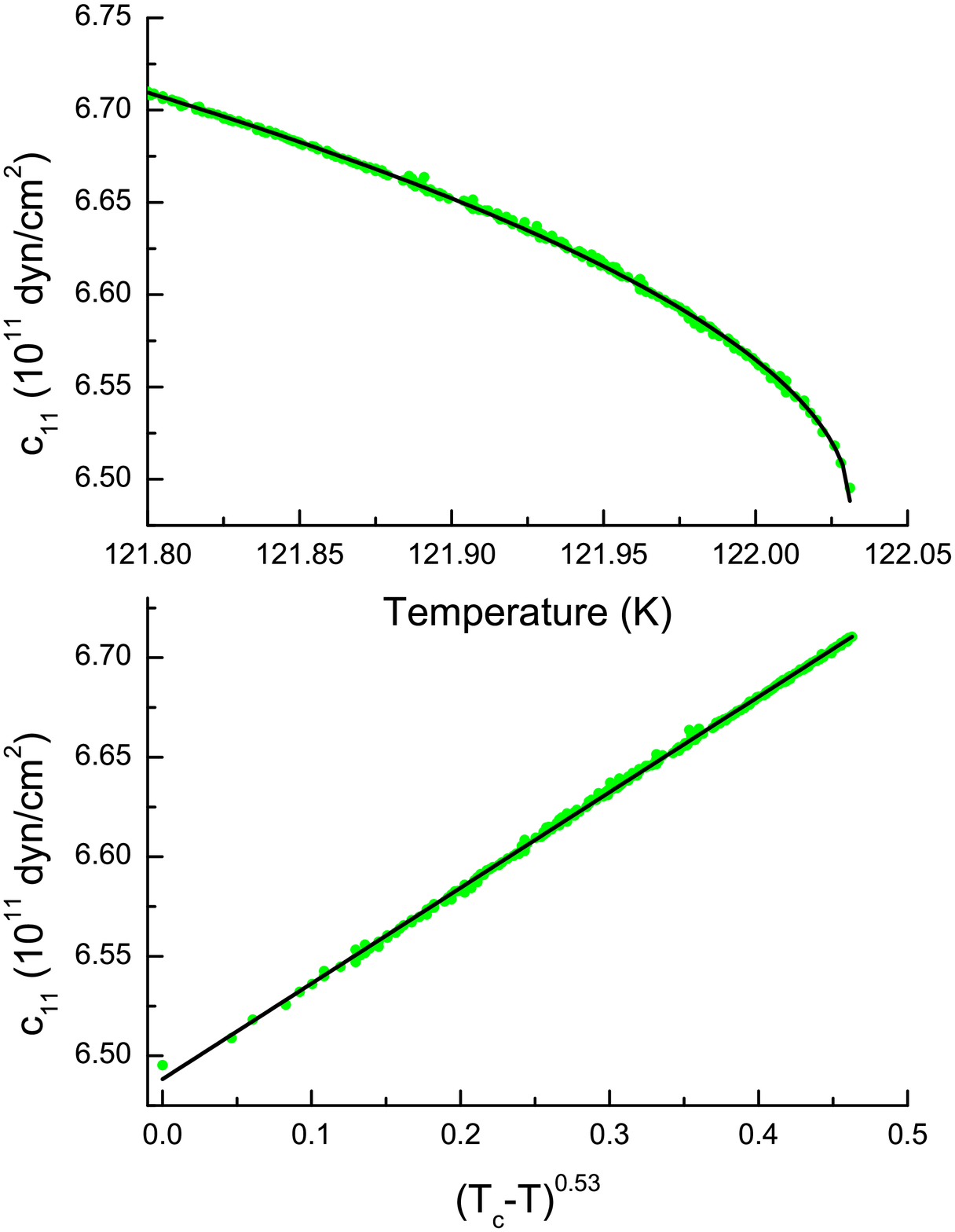}
\caption{\label{fig5} Temperature dependence of $c_{11}$ in the ferroelectric  phase of KDP.}
 \end{figure}
\begin{figure}[htb]
\includegraphics[width=75mm]{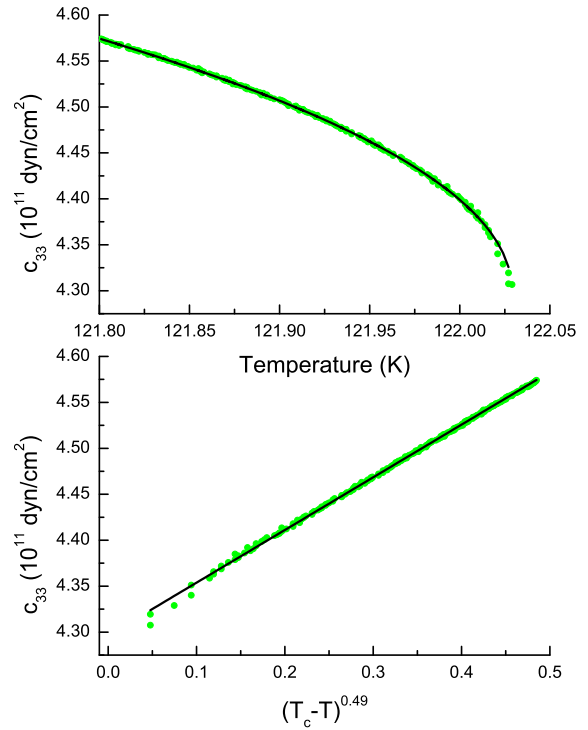}
\caption{\label{fig6} Temperature dependence of $c_{33}$ in the ferroelectric phase of KDP.}
 \end{figure}
\section{Results and discussion}
As is seen in Fig.~\ref{fig1}, the elastic moduli $c_{11}$ and $c_{33}$ both demonstrate strong softening when approaching the phase transition point from the ferroelectric phase. Their behavior can thus be described by a universal curve (see Fig.~\ref{fig2}). Both moduli are reduced by $\sim20\%$ when the temperature increases from $\sim$110 K to the phase transition point at 122 K. In contrast, as was pointed out earlier, the shear modulus $c_{66}$ shows softening both in the paraelectric and ferroelectric phases, satisfying the relation $c_{66}\sim(T-T_c)^{-1}$ at least in the paraelectric phase~\cite{Garland1969}. Finally, the evolution of the shear modulus $c_{44}$ across the phase transition looks absolutely flat, though it reveals some structure at higher magnification (the absolute fluctuation of $c_{44}$ in the entire temperature range studied does not exceed 1.5\%)
The first order nature of the phase transition is illustrated in Fig.~\ref{fig3}, \ref{fig4}, which show a small ($\sim$0.02K) but distinct hysteresis of the elastic properties of KDP across the transition.

The analysis of the temperature evolution of $c_{11}$ and $c_{33}$ in the ferroelectric phase shows that these quantities can be approximated by the expression $c_{ii}\sim(T-T_0)^{\alpha}$ with $\alpha\approx0.5$ in the temperature range 0.001K$<T_0-T<$0.2K (see Fig.~\ref{fig5}, \ref{fig6}). We note that our attempts to simultaneously determine the values of $\alpha$ and $T_0$ were not quite satisfactory. So, at first, we fixed $T_0$ at the (experimental) values of $T_c$, which are different for the cooling and heating runs, and found  $\alpha\approx0.5$. Then we fixed $\alpha$ at 0.5 and fitted $T_0$, which appeared (within the experimental resolution) not to be different from the experimental $T_c$ values. This implies that the temperature of the expected discontinuity of the elastic moduli is pretty close to the $T_0$ value in the relation $c_{ii}\sim(T-T_0)^{\alpha}$. This situation agrees completely with the presence of a well defined hysteresis. On the other hand it probably indicates that the transition temperature is very close to the absolute instability temperature. Remarkably, our analysis of the data in Ref.~\cite{Benepe1971, Bastie1975}  shows that in the restricted temperature interval near the phase transition point the order parameter exponent in KDP is close to  $0.5$. Whether this is a simple coincidence or results from some physical correlations remain to be seen. We recall now the suggestion that (see Ref.~\cite{Geguzina1967}) longitudinal elastic waves propagating in a piezo active media may locally change the polarization such that the velocity of longitudinal waves in KDP may not experience any kind of abnormalities at the phase transition.  The experimental data  obviously contradict this statement, which could probably be explained by the domain structure of real samples.

\section{Conclusions}
In conclusion, we report results of ultrasonic studies on a single crystal of KDP in the temperature range 2-300 K. The elastic moduli $c_{11}$, $c_{33}$, $c_{44}$ and $c_{66}$ were calculated from longitudinal and transverse sound velocities along the [001] and [100] directions. The analysis of the temperature evolution of $c_{11}$ and $c_{33}$ in the ferroelectric phase shows that these quantities can be approximated by the expression $c_{ii}\sim(T-T_0)^{0.5}$ in the temperature range 0.001K$<T_c-T<$0.2K for reasons that remain to be resolved. The elastic constants $c_{11}$, $c_{33}$ can be presented in a common reduced form, which again contradicts the conclusions made in Ref.~\cite{Geguzina1967}.

\section{Acknowledgment}
Technical help of A.N. Voronovskii is greatly appreciated. The Institute of Crystallography and Institute of Applied Physics of Russian Academy of Sciences provided the single crystals of KDP. We express special thanks to Mr. V.D. Loshkarev of Institute of Applied Physics for the sample preparation. We appreciate the support of the Russian Foundation for Basic Research (grant 12-02-00376-a), the Program of the Physics Department of RAS on Strongly Correlated Systems.

\end{document}